\documentclass[a4paper,fleqn]{cas-dc}
\usepackage[authoryear,longnamesfirst]{natbib}

\def\tsc#1{\csdef{#1}{\textsc{\lowercase{#1}}\xspace}}
\tsc{WGM}
\tsc{QE}

\begin{document}
\let\WriteBookmarks\relax
\def\floatpagepagefraction{1}
\def\textpagefraction{.001}

\shorttitle{Thermodynamics of charmed hadrons across chiral crossover  from lattice QCD}    
\shortauthors{S.~Sharma et al.}  
\title [mode = title]{Thermodynamics of charmed hadrons across chiral crossover  from lattice QCD}  
\author[1]{Sipaz Sharma}[orcid=0000-0001-6916-2233]
\ead{sipaz.sharma@tum.de}
\affiliation[1]{organization={Physik Department, Technische Universit\"at M\"unchen},
            addressline={James-Franck-Stra{\ss}e 1}, 
            city={Garching~b.~M\"unchen},
            postcode={D-85748}, 
            country={Germany}}
      
       \author[2]{Frithjof Karsch}
        \affiliation[2]{organization={Fakult\"at f\"ur Physik, Universit\"at Bielefeld},
        	addressline={Universitätsstraße 25}, 
        	city={Bielefeld},
        	postcode={D-33615}, 
        	country={Germany}}
          \author[3]{Peter Petreczky}
        \affiliation[3]{organization={Physics Department, Brookhaven National Laboratory},
        	city={Upton, NY},
        	postcode={11973}, 
        	country={USA}}

\begin{abstract}
We use up to fourth-order charm fluctuations and their correlations with net baryon number, electric charge, and strangeness fluctuations, calculated within the framework of lattice QCD, to study the continuum partial pressure contributions of charmed baryons and mesons. We show that, at and below the chiral crossover temperature, these partial pressures receive enhanced contributions from experimentally unobserved charmed hadrons predicted by the Quark Model. Additionally, we demonstrate that at the chiral crossover, the Hadron Resonance Gas description breaks down, signaling the modification of open charm hadrons and thereby implying the onset of charm deconfinement. We present evidence for the survival of low-lying non-radial $1S$ and $1P$ hadron-like excitations above the chiral crossover, which hints at the sequential melting of charmed hadrons. Finally, we investigate the continuum partial pressure contribution of charm quark-like excitation that emerges at the chiral crossover and calculate its temperature-dependent in-medium mass.
\end{abstract}

\ExplSyntaxOn
\keys_set:nn { stm / mktitle } { nologo }
\ExplSyntaxOff
\maketitle

\section{Introduction}
Lattice studies in the recent years have predicted the existence of experimentally unobserved charmed resonances by calculating charm fluctuations and their correlations with other conserved quantum numbers a.k.a. generalized charm susceptibilities \cite{BAZAVOV2014210, Sharma:2022ztl,  Bazavov:2023xzm}. Recently, analysis of experimental  open charm hadron yields using the extended Statistical Hadronization Model for charm (SHMc) \cite{ANDRONIC2019759}, showed enhanced yields in the charmed baryon sector \cite{Braun-Munzinger:2024ybd}. This enhancement is relative to the experimentally known charmed hadronic states tabulated in the Particle Data Group (PDG) records, and  therefore corroborates the existence of missing resonances in the open charm sector. 

Until now, lattice QCD predictions of missing resonances have been based solely on the ratios of generalized charm susceptibilities. This approach, however, could not provide a quantitative measure of enhancement in individual quantum number channels. The reason lattice QCD studies had to resort to ratios of generalized susceptibilities rather then the susceptibilities themselves is because
controlling the cutoff effects in the latter is challenging due to the relatively large charm quark mass.  In this study, for the first time, we use our continuum extrapolated lattice QCD results to explicitly determine enhancement factors at the chiral crossover temperature, ${T_{pc}=(156.5\pm1.5)}$ MeV \cite{HotQCD:2018pds, Borsanyi:2020fev}, for both the charmed baryonic and mesonic sectors.

The chiral crossover also exhibits deconfining feature in the charm quark sector, i.e. charm quarks as new degrees of freedom start appearing at or close to the pseudo-critical temperature of chiral symmetry restoration (Bazavov et al. (2024)). Clearly the appearance of charm quark degrees of freedom at $T_{pc}$, arising due to the disappearance of some hadronic states, will also liberate light quark degrees of freedom, although in the chiral limit this may not directly be related to the universal aspects of the chiral phase transition. The partial charm hadronic pressures start to deviate from the HRG model prediction, but not all charmed
hadrons dissociate at $T_{pc}$. Furthermore, recent lattice QCD studies provide compelling evidence for the survival of charmed hadrons above the chiral crossover temperature \cite{Bazavov:2023xzm, Sharma:2024ucs, Sharma:2024edf}. In this paper, we investigate the gradual deconfinement of charmed hadrons by analyzing continuum extrapolated proxies for partial pressures of charmed baryons and mesons across the chiral crossover.

\section{Calculating generalized susceptibilities on the lattice}
\label{sec: susc}
To project out the relevant degrees of freedom in the charm sector, one calculates the generalized susceptibilities, ${\chi^{BQSC}_{klmn}}$, of the conserved charges: baryon number, $B$, electric charge, $Q$, strangeness, $S$, and charm, $C$. This involves taking appropriate derivatives of the total pressure,
\begin{equation}
    P=\frac{T}{V}\ln Z(T,V,\hat{\mu}_B,\hat{\mu}_Q,\hat{\mu}_S,\hat{\mu}_C)
    \; ,
\label{pressure}
\end{equation}
which contains contributions from the total charm pressure, $P_C$. These derivatives are taken with respect to the chemical potentials, ${\hat{\mu}_X = \mu_X/T}$, ${\forall X \in \{B, Q, S, C\}}$, of the quantum number combinations one is interested in,
\begin{equation} 
	{\chi^{BQSC}_{klmn}=\dfrac{\partial^{(k+l+m+n)}\;\;[P\;(\hat{\mu}_B,\hat{\mu}_Q,\hat{\mu}_S,\hat{\mu}_C)\;/T^4]}{\partial\hat{\mu}^{k}_B\;\;\partial\hat{\mu}^{l}_Q\;\;\partial\hat{\mu}^{m}_S\;\;\partial\hat{\mu}^{n}_C}}\bigg|_{\overrightarrow{\mu}=0}\text{,}
	\label{eq:chi}
\end{equation}
where $\vec{\mu}=(\mu_B, \mu_Q, \mu_S, \mu_C)$. 
Note that ${\chi^{BQSC}_{klmn}}$ will be non-zero only for ${(k+l+m+n) \in \text{even}}$. In the following, if the subscript corresponding to a conserved charge is zero in the
left hand side of Eq. \eqref{eq:chi}, then both the corresponding superscript as well the zero subscript will be suppressed. 

Generalized susceptibilities up to fourth order have been calculated on the (2+1)-flavor HISQ (Highly Improved Staggered Quark) gauge configurations generated by the HotQCD collaboration \cite{Bollweg:2021vqf} for the physical strange to light quark mass ratio, ${m_s/m_{l}}=27$. The temperature scale was set using the $f_K$ scale \cite{Bollweg:2021vqf}. 
The charm-quark sector has been treated in the quenched approximation. We use HISQ action with epsilon term for charm quark to remove $\mathcal{O}((am_c)^4)$ tree-level lattice artifacts \cite{Follana:2006rc}. This setup is known to have small discretization effects for charm quark related observables \cite{Follana:2006rc,MILC:2010pul}. The calculation of $\chi^{BQSC}_{klmn}$ involves derivatives of the pressure and on the lattice this is achieved by the unbiased stochastic estimation of various traces -- consisting of inversions and derivatives  of the fermion matrices ($D$) -- using the random noise method \cite{Mitra:2022vtf}. 

Lattice cutoff effects cancel to a large extent in the ratios of various generalized susceptibilities  \cite{Bazavov:2023xzm}. These ratios, even though calculated on lattices with a finite lattice spacing, $a$, are close to
the continuum limit results.  Therefore, in this work we use generalized susceptibilities normalized by $\chi^C_4$ calculated on lattices with a fixed temporal extent, $N_\tau=8$, where $N_\tau=(aT)^{-1}$.   In order to understand the cut-off effects due to heavier charm quark, we used two different Lines of Constant Physics (LCPs) to tune the bare charm quark mass $am_c$.  The first LCP (LCP$_{[a]}$) corresponds to keeping the spin-averaged charmonium mass, ${(3m_{J/\psi}+m_{\eta_{c\bar{c}}})/4}$, fixed to its physical value \cite{Sharma:2022ztl}. The second LCP (LCP$_{[b]}$)  is defined by the physical (PDG) charm to strange quark mass ratio, $m_c/m_s=11.76$ \cite{ParticleDataGroup:2022pth}.  Further details of the charm-quark mass tuning and parametrization can be found in \cite {Sharma:2024ucs}. Here we use ratios of generalized susceptibilities calculated on lattices with temporal extent $N_\tau=8$ on LCP$_{[b]}$ because of their relatively smaller errors.  To convert generalized susceptibilities normalized by $\chi^C_4$ into absolute numbers we use continuum extrapolated results for $\chi^C_4$. Details of the continuum extrapolation will be given in a forthcoming  publication.

\section{Boltzmann Approximation}
\begin{figure*}[ht]
	
	\includegraphics[width=0.48\linewidth]{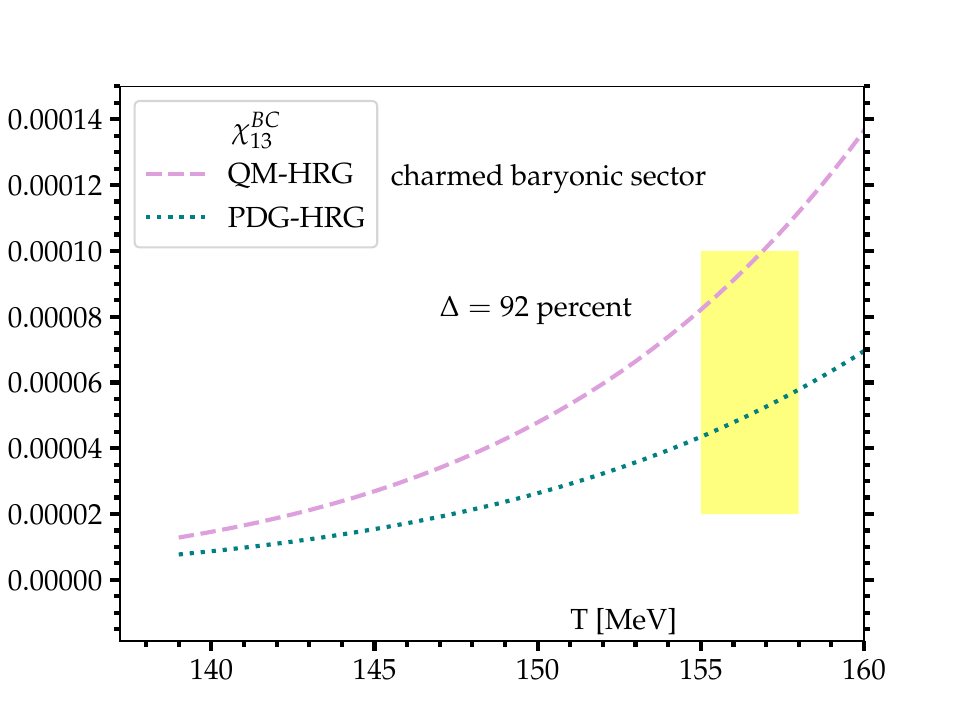}
	\includegraphics[width=0.48\linewidth]{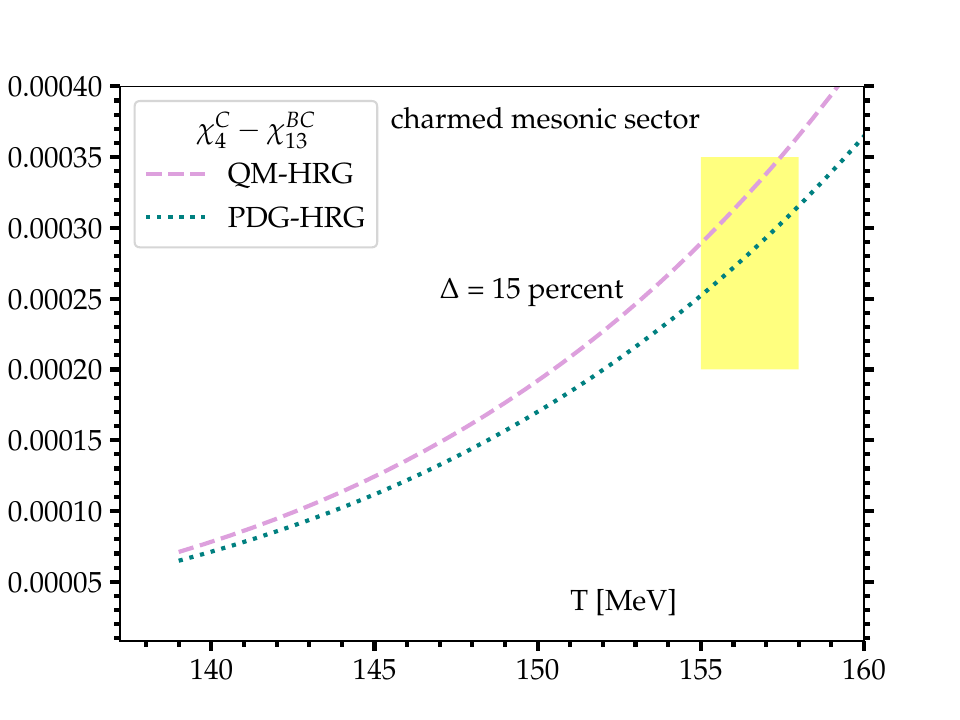}
	
	\caption{Comparison of QM-HRG and PDG-HRG predictions for charmed baryons [{\it Left}] and charmed mesons [{\it Right}]. $\Delta=100 |1-\text{QM-HRG}/\text{PDG-HRG}|_{T_{pc}}$. The yellow bands represent $T_{pc}$ with its uncertainty.}
	\label{fig:barCmesC}
\end{figure*}
\subsection{Charmed Hadrons}
Hadron resonance gas model (HRG) has been successful in describing the particle abundance ratios measured in the heavy-ion experiments. It describes a non-interacting gas of hadron resonances, and therefore can be used to calculate the hadronic pressure below ${T_{pc}}$ \cite{ANDRONIC2019759}. In the Boltzmann approximation, the dimensionless partial pressures from the charmed meson, ${P^C_{M}},$  and the charmed baryon, ${P^C_{B}}$, sectors take the following forms \cite{Allton:2005gk}:
\begin{gather}
	\begin{aligned}
		{P^C_{M}(T,\overrightarrow{\mu})}&{=\dfrac{1}{2\pi^2}\sum_{i\in \text{C-mesons}}g_i \bigg(\dfrac{m_i}{T}\bigg)^2K_2(m_i/T)}\\
			&{\text{cosh}(Q_i\hat{\mu}_Q+S_i\hat{\mu}_S+C_i\hat{\mu}_C)} \text{ ,}\\
		{P^C_{B}(T,\overrightarrow{\mu})}&={\dfrac{1}{2\pi^2}\sum_{i\in \text{C-baryons}}g_i \bigg(\dfrac{m_i}{T}\bigg)^2K_2(m_i/T)}\\
			&{\text{cosh}(B_i\hat{\mu}_B+Q_i\hat{\mu}_Q+S_i\hat{\mu}_S+C_i\hat{\mu}_C)} \text{ .}
		\label{eq:McBc}
	\end{aligned}
\end{gather}

In above equations, at a given temperature, the summation is over all charmed mesons/baryons (C-mesons/baryons) with masses given by ${m_i}$; degeneracy factors of the states with equal mass and same quantum numbers are represented by ${g_i}$; ${K_2(x)}$ is a modified Bessel function, which for a large argument can be approximated by
${K_2(x)}\sim\sqrt{\pi/2x}\;e^{-x}\;[1+\mathbb{O}(x^{-1})]$. Consequently, if a charmed state under consideration is much heavier than the relevant temperature scale, such that ${m_i\gg T}$, then the contribution to ${P_C}$ from that particular state will be exponentially suppressed, e.g., the singly-charmed
${\Lambda}_c^{+}$ baryon has a PDG mass of about $2286$~MeV, whereas the doubly-charmed ${\Xi_{cc}^{++}}$ baryon's mass as tabulated in PDG records is about $3621$ MeV, hence at ${T_{pc}}$, the contribution to ${P^C_{B}}$ from ${\Xi_{cc}^{++}}$ will be suppressed by a factor of $10^{-4}$ in relation to ${\Lambda}_c^{+}$ contribution. Therefore, in the validity regime of Boltzmann approximation, substituting Eq. \eqref{eq:McBc} in Eq. \eqref{eq:chi} implies $P_C(T,\vec{\mu})\approx\chi_2^C\approx\chi_n^C$, for $n$ even.

 Charm fluctuations calculated in the framework of lattice QCD can receive enhanced contributions due the existence of not-yet-discovered open-charm states. It is possible to compare this enhancement to the HRG calculations performed with two data sets. The first scenario, denoted by PDG-HRG, is based on the experimentally established states tabulated in the PDG records. The second scenario, denoted by QM-HRG, in addition to PDG states, takes into account states predicted via Quark-Model calculations \cite{Ebert:2009ua, Ebert:2011kk, Chen:2022asf}. In the validity regime of HRG, $\chi^{nm}_{BC}\approx P^C_{B}$, and $P^C_{M} = P_C-P^C_{B}$. Fig. \ref{fig:barCmesC} [{\it Left}] shows that the QM-HRG predicts a $92\%$ enhancement of the partial charmed baryon pressure at $T_{pc}$ compared to the PDG-HRG expectation, whereas the predicted enhancement in the partial charmed meson pressure is only $15\%$. These results suggest that, based on the QM-HRG predictions, the charmed baryonic sector is significantly more incomplete than the charmed mesonic sector. In recent years, the 
 experimental
 discovery of various $\Lambda_C$ and $\Omega_C$ resonances, in particular by LHCb and Belle collaborations \cite{Belle:2016tai,LHCb:2017uwr,LHCb:2017jym,Chen:2017gnu} have confirmed this.
\subsection{Charm Quarks}
Charm quarks offer an advantage over the light quarks because for temperatures a few times $T_{pc}$, the Boltzmann approximation works for an ideal massive quark-antiquark gas \cite{Allton:2005gk,BAZAVOV2014210}. Therefore, in this approximation, the dimensionless partial charm quark pressure, $P^C_{q}$, is given by, 

\begin{eqnarray}
	P^C_{q}(T,\overrightarrow{\mu})&=&\dfrac{3}{\pi^2}\bigg(\dfrac{m_q^C}{T}\bigg)^2K_2(m_q^C/T)\cdot \nonumber \\
	&&\text{cosh}\bigg(\dfrac{2}{3}\hat{\mu}_Q+\dfrac{1}{3}\hat{\mu}_B+\hat{\mu}_C\bigg) \text { ,}
	\label{eq:Qc}
\end{eqnarray}
where $m_q^C$ is the pole mass of the charm quark, and the degeneracy factor is 6.

\section{Enhancement of charmed hadrons}

\begin{figure*}[ht]
	
	\includegraphics[width=0.48\linewidth]{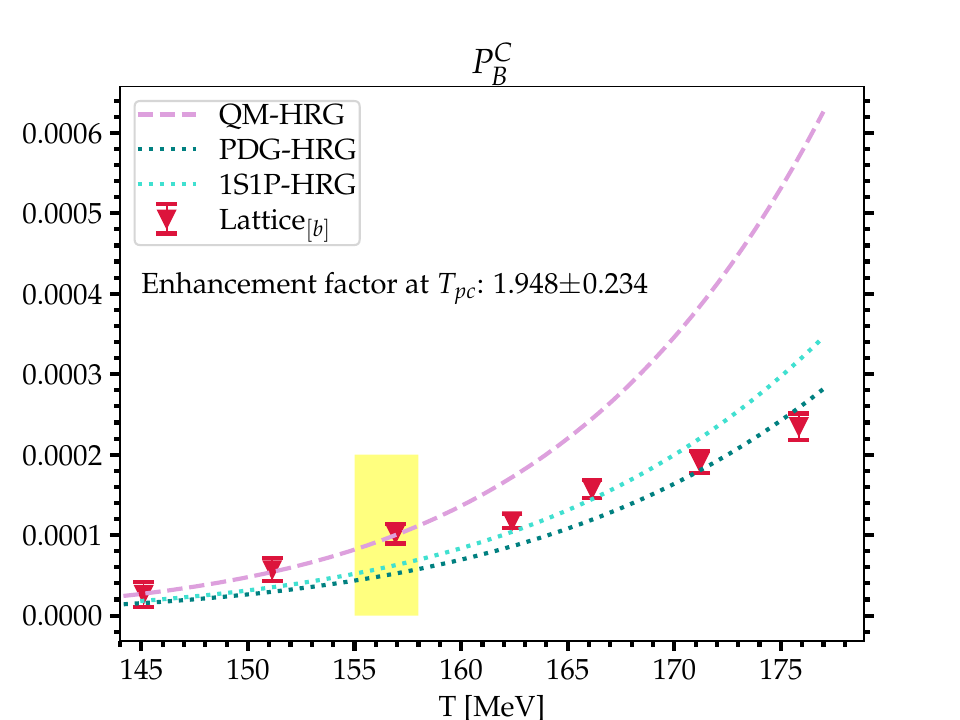}
	\includegraphics[width=0.48\linewidth]{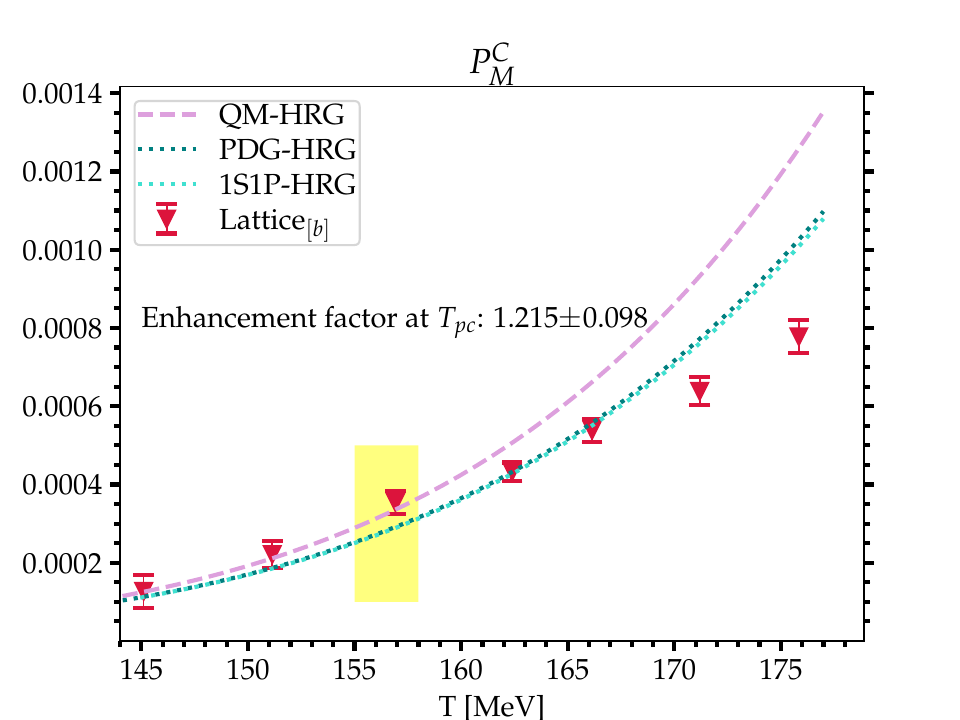}
	
	\caption{Shown are partial charmed baryon pressure [{\it Left}] and partial charmed meson pressure [{\it Right}] along with the respective QM-HRG, PDG-HRG and 1S1P-HRG (see text) predictions. The yellow bands represent $T_{pc}$ with its uncertainty.} 
	\label{fig:enhan}
\end{figure*}
To investigate the nature of charm degrees of freedom above  $T_{pc}$, we extend the simple hadron gas model allowing the presence of partial charm quark pressure based on Ref.~\cite{Mukherjee:2015mxc},
\begin{align}
	P_C(T,\vec{\mu})=P^C_{M}(T,\vec{\mu})+P^C_{B}(T,\vec{\mu})+P^C_{q}(T,\vec{\mu}) \, .
	\label{eq:Pmodel}
\end{align}
In our recent works \cite{Bazavov:2023xzm, Sharma:2024ucs, Sharma:2024edf}, we show that this model successfully passes numerous validity tests and satisfies various constraints \cite{Bazavov:2023xzm}. In this quasi-particle model, the partial pressures of quark-, baryon- and meson-like excitations for $\vec{\mu}=0$ can be expressed in terms of the generalized susceptibilities as follows,

\begin{align}
	P^C_{q}&=9(\chi^{BC}_{13}-\chi^{BC}_{22})/2\; , 
	\label{eq:partial-quasi} 	\\
	P^C_{B}&=(3\chi^{BC}_{22}-\chi^{BC}_{13})/2\; , 
	\label{eq:partial-quasiB}\\
	P^C_{M}&=\chi^{C}_{4}+3\chi^{BC}_{22}-4\chi^{BC}_{13} \; .
	\label{eq:quasi}
\end{align}
Note that in the validity regime of HRG, $\chi^{BC}_{22}\approx\chi^{BC}_{13}$. Hence, in this phase the partial pressure contribution from quark-like excitations, $P^C_q$ is, by construction, zero, and $P^C_B$ and $P^C_M$ reduce to $\chi^{BC}_{13}$ and $\chi^4_C-\chi^{BC}_{13}$, respectively.

Continuum estimates for charmed baryon and meson partial pressures shown in Fig. \ref{fig:enhan} are obtained by multiplying continuum extrapolated $\chi^C_4$ values to ratios $P^C_B/P_C$ and $P^C_M/P_C$  shown in our previous work \cite{Bazavov:2023xzm}. As mentioned earlier, these ratios, despite being calculated on a finite temporal lattice extent, are largely cut-off independent. For $T<T_{pc}$, Fig. \ref{fig:enhan} shows a clear agreement of the QM-HRG predictions with the  partial hadron pressures calculated on the lattice. This agreement, in addition to corroborating  the validity of HRG in the low temperature phase, confirms the existence of experimentally unobserved hadronic states. In particular, at the chiral crossover, in the charmed baryonic sector, $P^C_B$ calculated on the lattice is almost twice as large as the PDG-HRG expectation, whereas $P^C_M$ calculated on the lattice is around $20\%$ larger than the  PDG-HRG prediction. The enhancement factors explicitly quoted in Fig. \ref{fig:enhan} agree within errors with the enhancement predictions quoted in Fig. \ref{fig:barCmesC}. 

\section{Charm degrees of freedom in QGP}
\begin{figure*}[ht]
	\includegraphics[width=0.48\linewidth]{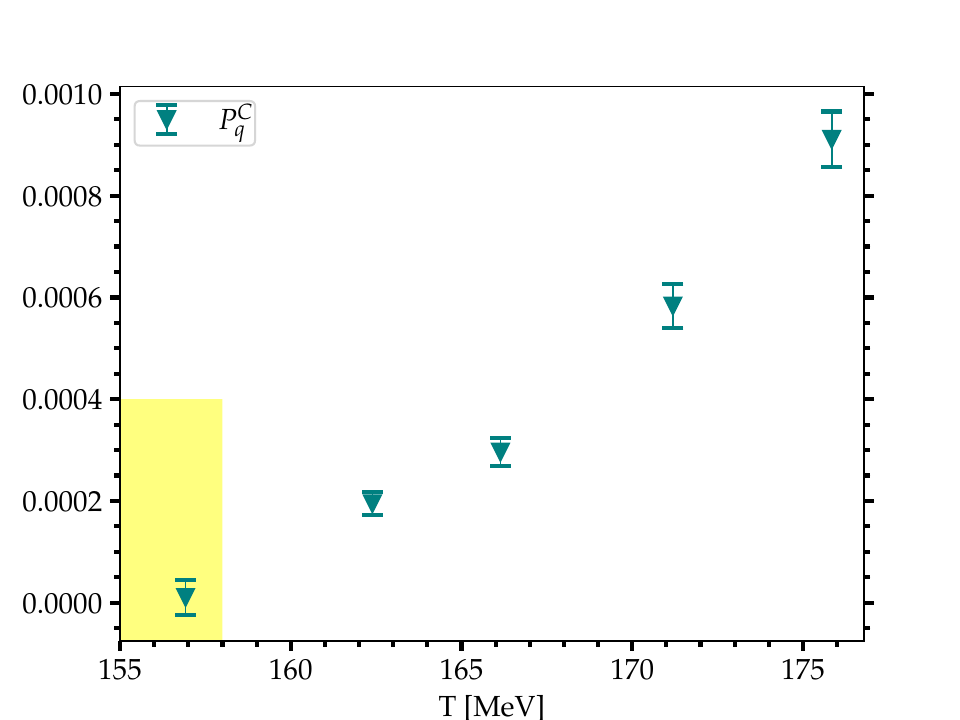}
		\includegraphics[width=0.48\linewidth]{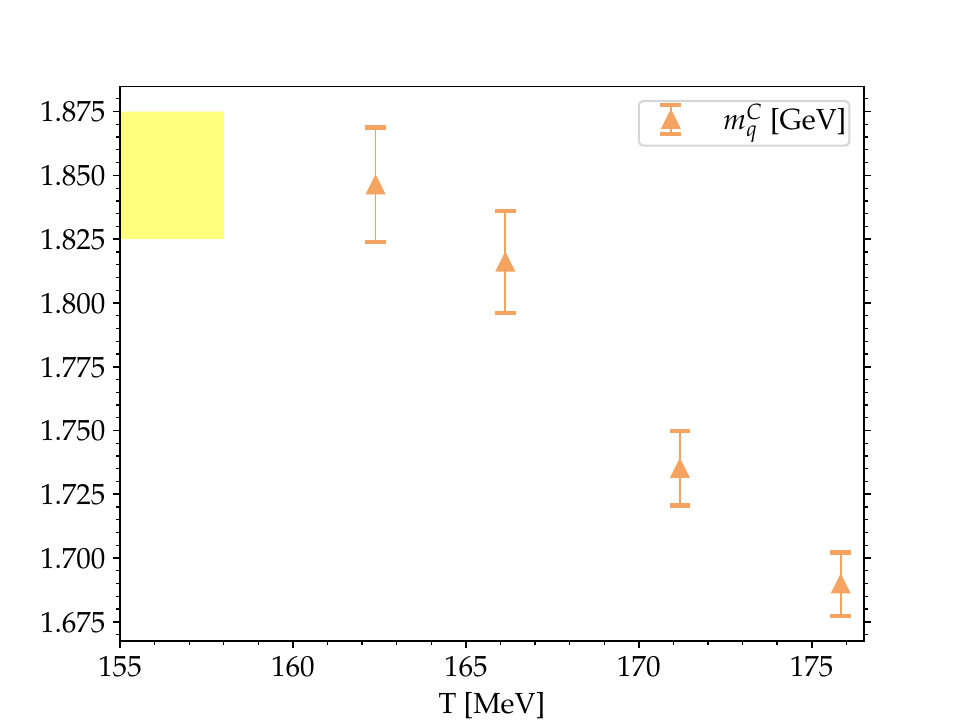}
	\caption{Shown is the partial pressure contribution of charm quark-like excitation above $T_{pc}$ [{\it Left}] and the temperature dependent in-medium mass of a charm quark-like excitation above $T_{pc}$ [{\it Right}]. The yellow bands represent $T_{pc}$ with its uncertainty.}
	\label{fig:Pq}
\end{figure*}
Fig. \ref{fig:Pq} [{\it Left}] shows that the partial pressure contribution to $P_C$ from charm quark-like excitation becomes non-zero at $T_{pc}$, and starts increasing as a function of temperature. This increasing $P^C_q$ results in a decreasing relative partial pressure contribution to $P_C$ from hadron-like excitations. At the highest temperature shown in Figs. \ref{fig:enhan} and \ref{fig:Pq} [{\it Left}], {\it i.e.,} $175$ MeV, the relative partial pressure contributions from hadron and quark-like excitations cross. Based on this trend, it is expected that above $T\sim175$ MeV, $P^C_q$ will become the dominant contribution. Therefore, eventually $P^C_B$ and $P^C_M$ shown in Fig. \ref{fig:enhan} will turn around and start decreasing. At sufficiently high  temperatures partial pressure contribution from charmed hadron-like excitations will go to zero.

Departure of charmed hadronic pressures from their respective QM-HRG predictions hints at a sequential melting pattern. This means that the higher excited charmed hadrons start dissociating at $T_{pc}$, whereas the ground state charmed hadrons survive inside the QGP. To put this on more solid grounds, we performed another HRG model calculation based on the low-lying, $1S$ and $1P$, charmed hadrons tabulated in \cite{Chen:2022asf} -- we label this as 1S1P-HRG. For $T_{pc}<T\leq 166.1$~MeV, both $P^C_B$ and $P^C_M$ in Fig. \ref{fig:enhan} are described by 1S1P-HRG. For $P^C_M$, both PDG-HRG and 1S1P-HRG almost overlap, and only at the highest temperatures shown in Fig. \ref{fig:enhan} [{\it Right}] there are some visible differences. This is expected because all low-lying charmed meson states are experimentally known. In addition to low-lying 1S and 1P states, PDG-HRG contains one charmed-light meson: $D_3^*(2750)$ and two charmed-strange mesons: $D_{s1}^*(2700)$ and $D_{s3}^*(2860)$. We used PDG masses of $D$, $D^*$, $D^*_0$ and $D_1$ charmed-light mesons. If chiral partners were to get degenerate at $T_{pc}$, then the scalar and axial-vector mesons would get lighter, 
making 1S1P-HRG curve overshoot $P^C_M$ above $T_{pc}$. This indicates that the charm sector is not strongly influenced 
by the chiral symmetry restoration. This is also in line with studies of screening masses for charmed parity partners, which indicate that unlike in the light quark sector the charmed screening masses  become almost degenerate only at temperatures above $2T_{pc}$ \cite{Bazavov:2014cta}. Similar observation holds for the charmed-strange mesonic sector. In the charmed baryonic sector, after departure from 1S1P-HRG, for the highest two temperatures, PDG-HRG prediction can very well describe $P^C_B$ in Fig. \ref{fig:enhan} [{\it Left}]. The fact that 1S1P-HRG lies above PDG-HRG for $P^C_B$ indicates that there are many low-lying and therefore thermodynamically dominant charmed baryons missing from the PDG record.

In the coexistence phase of charmed hadron and quark-like excitations, $m^C_q$ in Eq. \eqref{eq:Qc} becomes temperature dependent and can be interpreted as the mass of a quasi-particle with quantum numbers of charm quark. Given $P^C_q(T)$ as a function of temperature, one can solve Eq. \eqref{eq:Qc} at each temperature and obtain temperature dependence of $m^C_q$. In Fig. \ref{fig:Pq} [{\it Right}], at each temperature, the error on $m^C_q$ is the standard deviation of $m^C_q$ values obtained after solving Eq.~\eqref{eq:Qc} for $50$ fake Gaussian samples. Fig. \ref{fig:Pq} [{\it Right}] shows that at $T=162.4$ MeV, $m^C_q$ is around the mass of D-meson, and starts decreasing with temperature. Quasi-particle model in Eq. \eqref{eq:Pmodel} considers a non-interacting gas of charmed hadron and quark-like excitations, but the temperature dependence of $m^C_q$ encodes the in-medium interactions. In the non-interacting quark gas limit, $m^C_q$ will become the pole mass of charm quark.

\section{Conclusions}
In this work, for the first time, we showed that in the low temperature phase HRG calculations agree with various continuum extrapolated generalized susceptibilities of charm calculated in the framework of lattice QCD when adding
experimentally unobserved charmed hadrons, predicted by the Quark Model calculations, to the HRG model spectrum. Our previous studies, predicting the existence of missing charmed resonances, utilised ratios of various generalized susceptibilities. We showed that continuum extrapolated partial pressures of charmed baryons and mesons calculated on the lattice receive enhanced contributions from these missing resonances. We concluded that the charmed baryon sector is significantly more incomplete than the charmed meson sector.

We showed that above chiral crossover, the charm pressure can be decomposed into partial pressures of charm quark, charmed mesons, and charmed baryon-like excitations. At the highest temperature, $T \sim 176$ MeV, studied by us, the charm pressure receives equal contributions from charmed hadrons and charm quark, indicating that charmed hadrons can exist in Quark Gluon Plasma (QGP).
\section*{Acknowledgments}

This work was supported by The Deutsche Forschungsgemeinschaft (DFG, German Research Foundation) - Project number 315477589-TRR 211,
”Strong interaction matter under extreme conditions”. This material is based upon work supported by The U.S. Department of Energy, Office of Science, Office of Nuclear Physics through Contract No.~DE-SC0012704 and the Topical Collaboration in Nuclear Theory \textit{Heavy-Flavor Theory (HEFTY) for QCD Matter}.
The authors gratefully acknowledge the
computing time and support provided to them on the high-performance computer Noctua 2 at the NHR Center
PC2 under the project name: hpc-prf-cfpd. These are funded by the Federal Ministry of Education
and Research and the state governments participating on the basis of the resolutions of the GWK
for the national high-performance computing at universities (www.nhr-verein.de/unsere-partner).
Numerical calculations have also been performed on the
GPU-cluster at Bielefeld University, Germany. We thank the Bielefeld HPC.NRW team for their support. 

All computations in this work were performed using \texttt{SIMULATeQCD} code ~\cite{HotQCD:2023ghu}. All the HRG calculations were performed using the AnalysisToolbox code developed by the HotQCD Collaboration \cite{Clarke:2023sfy}.

\bibliographystyle{cas-model2-names}
\bibliography{refs}

\begin{thebibliography}{26}
\expandafter\ifx\csname natexlab\endcsname\relax\def\natexlab#1{#1}\fi
\providecommand{\url}[1]{\texttt{#1}}
\providecommand{\href}[2]{#2}
\providecommand{\path}[1]{#1}
\providecommand{\DOIprefix}{doi:}
\providecommand{\ArXivprefix}{arXiv:}
\providecommand{\URLprefix}{URL: }
\providecommand{\Pubmedprefix}{pmid:}
\providecommand{\doi}[1]{\href{http://dx.doi.org/#1}{\path{#1}}}
\providecommand{\Pubmed}[1]{\href{pmid:#1}{\path{#1}}}
\providecommand{\bibinfo}[2]{#2}
\ifx\xfnm\relax \def\xfnm[#1]{\unskip,\space#1}\fi
\bibitem[{Aaij et~al.(2017a)}]{LHCb:2017uwr}
\bibinfo{author}{Aaij, R.}, et~al. (\bibinfo{collaboration}{LHCb}),
  \bibinfo{year}{2017}a.
\newblock \bibinfo{title}{{Observation of five new narrow $\Omega_c^0$ states
  decaying to $\Xi_c^+ K^-$}}.
\newblock \bibinfo{journal}{Phys. Rev. Lett.} \bibinfo{volume}{118},
  \bibinfo{pages}{182001}.
\newblock \DOIprefix\doi{10.1103/PhysRevLett.118.182001},
  \href{http://arxiv.org/abs/1703.04639}{\tt arXiv:1703.04639}.
\bibitem[{Aaij et~al.(2017b)}]{LHCb:2017jym}
\bibinfo{author}{Aaij, R.}, et~al. (\bibinfo{collaboration}{LHCb}),
  \bibinfo{year}{2017}b.
\newblock \bibinfo{title}{{Study of the $D^0 p$ amplitude in $\Lambda_b^0\to
  D^0 p \pi^-$ decays}}.
\newblock \bibinfo{journal}{JHEP} \bibinfo{volume}{05}, \bibinfo{pages}{030}.
\newblock \DOIprefix\doi{10.1007/JHEP05(2017)030},
  \href{http://arxiv.org/abs/1701.07873}{\tt arXiv:1701.07873}.
\bibitem[{Allton et~al.(2005)Allton, Doring, Ejiri, Hands, Kaczmarek, Karsch,
  Laermann and Redlich}]{Allton:2005gk}
\bibinfo{author}{Allton, C.R.}, \bibinfo{author}{Doring, M.},
  \bibinfo{author}{Ejiri, S.}, \bibinfo{author}{Hands, S.J.},
  \bibinfo{author}{Kaczmarek, O.}, \bibinfo{author}{Karsch, F.},
  \bibinfo{author}{Laermann, E.}, \bibinfo{author}{Redlich, K.},
  \bibinfo{year}{2005}.
\newblock \bibinfo{title}{{Thermodynamics of two flavor QCD to sixth order in
  quark chemical potential}} \bibinfo{volume}{71}, \bibinfo{pages}{054508}.
\newblock \DOIprefix\doi{10.1103/PhysRevD.71.054508},
  \href{http://arxiv.org/abs/hep-lat/0501030}{\tt arXiv:hep-lat/0501030}.
\bibitem[{Andronic et~al.(2019)Andronic, Braun-Munzinger, K\"ohler and
  Stachel}]{ANDRONIC2019759}
\bibinfo{author}{Andronic, A.}, \bibinfo{author}{Braun-Munzinger, P.},
  \bibinfo{author}{K\"ohler, M.}, \bibinfo{author}{Stachel, J.},
  \bibinfo{year}{2019}.
\newblock \bibinfo{title}{Testing charm quark thermalisation within the
  statistical hadronisation model}.
\newblock \bibinfo{journal}{Nuclear Physics A} \bibinfo{volume}{982},
  \bibinfo{pages}{759--762}.
\newblock \URLprefix
  \url{https://www.sciencedirect.com/science/article/pii/S0375947418301921},
  \DOIprefix\doi{https://doi.org/10.1016/j.nuclphysa.2018.09.004}.
  \bibinfo{note}{the 27th International Conference on Ultrarelativistic
  Nucleus-Nucleus Collisions: Quark Matter 2018}.
\bibitem[{Bazavov et~al.(2024)Bazavov, Bollweg, Kaczmarek, Karsch, Mukherjee,
  Petreczky, Schmidt and Sharma}]{Bazavov:2023xzm}
\bibinfo{author}{Bazavov, A.}, \bibinfo{author}{Bollweg, D.},
  \bibinfo{author}{Kaczmarek, O.}, \bibinfo{author}{Karsch, F.},
  \bibinfo{author}{Mukherjee, S.}, \bibinfo{author}{Petreczky, P.},
  \bibinfo{author}{Schmidt, C.}, \bibinfo{author}{Sharma, S.},
  \bibinfo{year}{2024}.
\newblock \bibinfo{title}{{Charm degrees of freedom in hot matter from lattice
  QCD}}.
\newblock \bibinfo{journal}{Phys. Lett. B} \bibinfo{volume}{850},
  \bibinfo{pages}{138520}.
\newblock \DOIprefix\doi{10.1016/j.physletb.2024.138520},
  \href{http://arxiv.org/abs/2312.12857}{\tt arXiv:2312.12857}.
\bibitem[{Bazavov et~al.(2014)Bazavov, Ding, Hegde, Kaczmarek, Karsch,
  Laermann, Maezawa, Mukherjee, Ohno, Petreczky, Schmidt, Sharma, Soeldner and
  Wagner}]{BAZAVOV2014210}
\bibinfo{author}{Bazavov, A.}, \bibinfo{author}{Ding, H.T.},
  \bibinfo{author}{Hegde, P.}, \bibinfo{author}{Kaczmarek, O.},
  \bibinfo{author}{Karsch, F.}, \bibinfo{author}{Laermann, E.},
  \bibinfo{author}{Maezawa, Y.}, \bibinfo{author}{Mukherjee, S.},
  \bibinfo{author}{Ohno, H.}, \bibinfo{author}{Petreczky, P.},
  \bibinfo{author}{Schmidt, C.}, \bibinfo{author}{Sharma, S.},
  \bibinfo{author}{Soeldner, W.}, \bibinfo{author}{Wagner, M.},
  \bibinfo{year}{2014}.
\newblock \bibinfo{title}{The melting and abundance of open charm hadrons}.
\newblock \bibinfo{journal}{Physics Letters B} \bibinfo{volume}{737},
  \bibinfo{pages}{210--215}.
\newblock \URLprefix
  \url{https://www.sciencedirect.com/science/article/pii/S0370269314006017},
  \DOIprefix\doi{https://doi.org/10.1016/j.physletb.2014.08.034}.
\bibitem[{Bazavov et~al.(2015)Bazavov, Karsch, Maezawa, Mukherjee and
  Petreczky}]{Bazavov:2014cta}
\bibinfo{author}{Bazavov, A.}, \bibinfo{author}{Karsch, F.},
  \bibinfo{author}{Maezawa, Y.}, \bibinfo{author}{Mukherjee, S.},
  \bibinfo{author}{Petreczky, P.}, \bibinfo{year}{2015}.
\newblock \bibinfo{title}{{In-medium modifications of open and hidden
  strange-charm mesons from spatial correlation functions}}.
\newblock \bibinfo{journal}{Phys. Rev. D} \bibinfo{volume}{91},
  \bibinfo{pages}{054503}.
\newblock \DOIprefix\doi{10.1103/PhysRevD.91.054503},
  \href{http://arxiv.org/abs/1411.3018}{\tt arXiv:1411.3018}.
\bibitem[{Bazavov et~al.(2010)}]{MILC:2010pul}
\bibinfo{author}{Bazavov, A.}, et~al. (\bibinfo{collaboration}{MILC}),
  \bibinfo{year}{2010}.
\newblock \bibinfo{title}{{Scaling studies of QCD with the dynamical HISQ
  action}}.
\newblock \bibinfo{journal}{Phys. Rev. D} \bibinfo{volume}{82},
  \bibinfo{pages}{074501}.
\newblock \DOIprefix\doi{10.1103/PhysRevD.82.074501},
  \href{http://arxiv.org/abs/1004.0342}{\tt arXiv:1004.0342}.
\bibitem[{Bazavov et~al.(2019)}]{HotQCD:2018pds}
\bibinfo{author}{Bazavov, A.}, et~al. (\bibinfo{collaboration}{HotQCD}),
  \bibinfo{year}{2019}.
\newblock \bibinfo{title}{{Chiral crossover in QCD at zero and non-zero
  chemical potentials}}.
\newblock \bibinfo{journal}{Phys. Lett. B} \bibinfo{volume}{795},
  \bibinfo{pages}{15--21}.
\newblock \DOIprefix\doi{10.1016/j.physletb.2019.05.013},
  \href{http://arxiv.org/abs/1812.08235}{\tt arXiv:1812.08235}.
\bibitem[{Bollweg et~al.(2021)Bollweg, Goswami, Kaczmarek, Karsch, Mukherjee,
  Petreczky, Schmidt and Scior}]{Bollweg:2021vqf}
\bibinfo{author}{Bollweg, D.}, \bibinfo{author}{Goswami, J.},
  \bibinfo{author}{Kaczmarek, O.}, \bibinfo{author}{Karsch, F.},
  \bibinfo{author}{Mukherjee, S.}, \bibinfo{author}{Petreczky, P.},
  \bibinfo{author}{Schmidt, C.}, \bibinfo{author}{Scior, P.}
  (\bibinfo{collaboration}{HotQCD}), \bibinfo{year}{2021}.
\newblock \bibinfo{title}{{Second order cumulants of conserved charge
  fluctuations revisited: Vanishing chemical potentials}}.
\newblock \bibinfo{journal}{Phys. Rev. D} \bibinfo{volume}{104}.
\newblock \DOIprefix\doi{10.1103/PhysRevD.104.074512},
  \href{http://arxiv.org/abs/2107.10011}{\tt arXiv:2107.10011}.
\bibitem[{Borsanyi et~al.()Borsanyi, Fodor, Guenther, Kara, Katz, Parotto,
  Pasztor, Ratti and Szabo}]{Borsanyi:2020fev}
\bibinfo{author}{Borsanyi, S.}, \bibinfo{author}{Fodor, Z.},
  \bibinfo{author}{Guenther, J.N.}, \bibinfo{author}{Kara, R.},
  \bibinfo{author}{Katz, S.D.}, \bibinfo{author}{Parotto, P.},
  \bibinfo{author}{Pasztor, A.}, \bibinfo{author}{Ratti, C.},
  \bibinfo{author}{Szabo, K.K.}, .
\newblock \bibinfo{title}{{QCD Crossover at Finite Chemical Potential from
  Lattice Simulations}} \bibinfo{volume}{125}, \bibinfo{pages}{052001}.
\newblock \DOIprefix\doi{10.1103/PhysRevLett.125.052001},
  \href{http://arxiv.org/abs/2002.02821}{\tt arXiv:2002.02821}.
\bibitem[{Braun-Munzinger et~al.(2024)Braun-Munzinger, Redlich, Sharma and
  Stachel}]{Braun-Munzinger:2024ybd}
\bibinfo{author}{Braun-Munzinger, P.}, \bibinfo{author}{Redlich, K.},
  \bibinfo{author}{Sharma, N.}, \bibinfo{author}{Stachel, J.},
  \bibinfo{year}{2024}.
\newblock \bibinfo{title}{{Emergence of New Systematics for Open Charm
  Production in High Energy Collisions}}
  \href{http://arxiv.org/abs/2408.07496}{\tt arXiv:2408.07496}.
\bibitem[{Chen and Liu(2017)}]{Chen:2017gnu}
\bibinfo{author}{Chen, B.}, \bibinfo{author}{Liu, X.}, \bibinfo{year}{2017}.
\newblock \bibinfo{title}{{New $\Omega_c^0$ baryons discovered by LHCb as the
  members of $1P$ and $2S$ states}}.
\newblock \bibinfo{journal}{Phys. Rev. D} \bibinfo{volume}{96},
  \bibinfo{pages}{094015}.
\newblock \DOIprefix\doi{10.1103/PhysRevD.96.094015},
  \href{http://arxiv.org/abs/1704.02583}{\tt arXiv:1704.02583}.
\bibitem[{Chen et~al.(2023)Chen, Chen, Liu, Liu and Zhu}]{Chen:2022asf}
\bibinfo{author}{Chen, H.X.}, \bibinfo{author}{Chen, W.}, \bibinfo{author}{Liu,
  X.}, \bibinfo{author}{Liu, Y.R.}, \bibinfo{author}{Zhu, S.L.},
  \bibinfo{year}{2023}.
\newblock \bibinfo{title}{{An updated review of the new hadron states}}.
\newblock \bibinfo{journal}{Rept. Prog. Phys.} \bibinfo{volume}{86},
  \bibinfo{pages}{026201}.
\newblock \DOIprefix\doi{10.1088/1361-6633/aca3b6},
  \href{http://arxiv.org/abs/2204.02649}{\tt arXiv:2204.02649}.
\bibitem[{Clarke et~al.(2024)Clarke, Altenkort, Goswami and
  Sandmeyer}]{Clarke:2023sfy}
\bibinfo{author}{Clarke, D.A.}, \bibinfo{author}{Altenkort, L.},
  \bibinfo{author}{Goswami, J.}, \bibinfo{author}{Sandmeyer, H.},
  \bibinfo{year}{2024}.
\newblock \bibinfo{title}{{Streamlined data analysis in Python}}.
\newblock \bibinfo{journal}{PoS} \bibinfo{volume}{LATTICE2023},
  \bibinfo{pages}{136}.
\newblock \DOIprefix\doi{10.22323/1.453.0136},
  \href{http://arxiv.org/abs/2308.06652}{\tt arXiv:2308.06652}.
\bibitem[{Ebert et~al.(2010)Ebert, Faustov and Galkin}]{Ebert:2009ua}
\bibinfo{author}{Ebert, D.}, \bibinfo{author}{Faustov, R.N.},
  \bibinfo{author}{Galkin, V.O.}, \bibinfo{year}{2010}.
\newblock \bibinfo{title}{{Heavy-light meson spectroscopy and Regge
  trajectories in the relativistic quark model}}.
\newblock \bibinfo{journal}{Eur. Phys. J. C} \bibinfo{volume}{66},
  \bibinfo{pages}{197--206}.
\newblock \DOIprefix\doi{10.1140/epjc/s10052-010-1233-6},
  \href{http://arxiv.org/abs/0910.5612}{\tt arXiv:0910.5612}.
\bibitem[{Ebert et~al.(2011)Ebert, Faustov and Galkin}]{Ebert:2011kk}
\bibinfo{author}{Ebert, D.}, \bibinfo{author}{Faustov, R.N.},
  \bibinfo{author}{Galkin, V.O.}, \bibinfo{year}{2011}.
\newblock \bibinfo{title}{{Spectroscopy and Regge trajectories of heavy baryons
  in the relativistic quark-diquark picture}}.
\newblock \bibinfo{journal}{Phys. Rev. D} \bibinfo{volume}{84},
  \bibinfo{pages}{014025}.
\newblock \DOIprefix\doi{10.1103/PhysRevD.84.014025},
  \href{http://arxiv.org/abs/1105.0583}{\tt arXiv:1105.0583}.
\bibitem[{Follana et~al.(2007)Follana, Mason, Davies, Hornbostel, Lepage,
  Shigemitsu, Trottier and Wong}]{Follana:2006rc}
\bibinfo{author}{Follana, E.}, \bibinfo{author}{Mason, Q.},
  \bibinfo{author}{Davies, C.}, \bibinfo{author}{Hornbostel, K.},
  \bibinfo{author}{Lepage, G.}, \bibinfo{author}{Shigemitsu, J.},
  \bibinfo{author}{Trottier, H.}, \bibinfo{author}{Wong, K.}
  (\bibinfo{collaboration}{HPQCD, UKQCD}), \bibinfo{year}{2007}.
\newblock \bibinfo{title}{{Highly improved staggered quarks on the lattice,
  with applications to charm physics}}.
\newblock \bibinfo{journal}{Phys. Rev. D} \bibinfo{volume}{75},
  \bibinfo{pages}{054502}.
\newblock \DOIprefix\doi{10.1103/PhysRevD.75.054502},
  \href{http://arxiv.org/abs/hep-lat/0610092}{\tt arXiv:hep-lat/0610092}.
\bibitem[{Kato et~al.(2016)}]{Belle:2016tai}
\bibinfo{author}{Kato, Y.}, et~al. (\bibinfo{collaboration}{Belle}),
  \bibinfo{year}{2016}.
\newblock \bibinfo{title}{{Studies of charmed strange baryons in the $\Lambda$D
  final state at Belle}}.
\newblock \bibinfo{journal}{Phys. Rev. D} \bibinfo{volume}{94},
  \bibinfo{pages}{032002}.
\newblock \DOIprefix\doi{10.1103/PhysRevD.94.032002},
  \href{http://arxiv.org/abs/1605.09103}{\tt arXiv:1605.09103}.
\bibitem[{Mazur et~al.(2023)}]{HotQCD:2023ghu}
\bibinfo{author}{Mazur, L.}, et~al. (\bibinfo{collaboration}{HotQCD}),
  \bibinfo{year}{2023}.
\newblock \bibinfo{title}{{SIMULATeQCD: A simple multi-GPU lattice code for QCD
  calculations}} \href{http://arxiv.org/abs/2306.01098}{\tt arXiv:2306.01098}.
\bibitem[{Mitra et~al.(2022)Mitra, Hegde and Schmidt}]{Mitra:2022vtf}
\bibinfo{author}{Mitra, S.}, \bibinfo{author}{Hegde, P.},
  \bibinfo{author}{Schmidt, C.}, \bibinfo{year}{2022}.
\newblock \bibinfo{title}{{New way to resum the lattice QCD Taylor series
  equation of state at finite chemical potential}}.
\newblock \bibinfo{journal}{Phys. Rev. D} \bibinfo{volume}{106},
  \bibinfo{pages}{034504}.
\newblock \DOIprefix\doi{10.1103/PhysRevD.106.034504},
  \href{http://arxiv.org/abs/2205.08517}{\tt arXiv:2205.08517}.
\bibitem[{Mukherjee et~al.(2016)Mukherjee, Petreczky and
  Sharma}]{Mukherjee:2015mxc}
\bibinfo{author}{Mukherjee, S.}, \bibinfo{author}{Petreczky, P.},
  \bibinfo{author}{Sharma, S.}, \bibinfo{year}{2016}.
\newblock \bibinfo{title}{{Charm degrees of freedom in the quark gluon
  plasma}}.
\newblock \bibinfo{journal}{Phys. Rev. D} \bibinfo{volume}{93},
  \bibinfo{pages}{014502}.
\newblock \DOIprefix\doi{10.1103/PhysRevD.93.014502},
  \href{http://arxiv.org/abs/1509.08887}{\tt arXiv:1509.08887}.
\bibitem[{Sharma(2023)}]{Sharma:2022ztl}
\bibinfo{author}{Sharma, S.}, \bibinfo{year}{2023}.
\newblock \bibinfo{title}{{Charm fluctuations in (2+1)-flavor QCD at high
  temperature}}.
\newblock \bibinfo{journal}{PoS} \bibinfo{volume}{LATTICE2022},
  \bibinfo{pages}{191}.
\newblock \DOIprefix\doi{10.22323/1.430.0191},
  \href{http://arxiv.org/abs/2212.11148}{\tt arXiv:2212.11148}.
\bibitem[{Sharma(2024a)}]{Sharma:2024ucs}
\bibinfo{author}{Sharma, S.} (\bibinfo{collaboration}{HotQCD}),
  \bibinfo{year}{2024}a.
\newblock \bibinfo{title}{{Charm Fluctuations and Deconfinement}}.
\newblock \bibinfo{journal}{PoS} \bibinfo{volume}{LATTICE2023},
  \bibinfo{pages}{200}.
\newblock \DOIprefix\doi{10.22323/1.453.0200},
  \href{http://arxiv.org/abs/2401.01194}{\tt arXiv:2401.01194}.
\bibitem[{Sharma(2024b)}]{Sharma:2024edf}
\bibinfo{author}{Sharma, S.}, \bibinfo{year}{2024}b.
\newblock \bibinfo{title}{{Persistence of charmed hadrons in QGP from lattice
  QCD}}, in: \bibinfo{booktitle}{{23rd Zimanyi School Winter Workshop}}.
\newblock \href{http://arxiv.org/abs/2410.04222}{\tt arXiv:2410.04222}.
\bibitem[{Workman et~al.(2022)}]{ParticleDataGroup:2022pth}
\bibinfo{author}{Workman, R.L.}, et~al. (\bibinfo{collaboration}{Particle Data
  Group}), \bibinfo{year}{2022}.
\newblock \bibinfo{title}{{Review of Particle Physics}}.
\newblock \bibinfo{journal}{PTEP} \bibinfo{volume}{2022},
  \bibinfo{pages}{083C01}.
\newblock \DOIprefix\doi{10.1093/ptep/ptac097}.

\end{thebibliography}

\end{document}